\documentclass[%
pra%
,reprint%
,twocolumn%
,secnumarabic%
,floatfix%
,superscriptaddress, showpacs%
,amssymb, amsmath, nobibnotes, aps]{revtex4-1}

\usepackage[latin1]{inputenc}
\usepackage{epsfig}
\usepackage{color}

\newcommand{\ind}[1]{_{\text{#1}}}
\newcommand{\bra}[1]{\left<#1\right|}
\newcommand{\ket}[1]{\left|#1\right>}

\begin{document}

\title{Non-linear absorption and density dependent dephasing in Rydberg EIT-media}
\author{Martin G\"arttner}
\affiliation{Max-Planck-Institut f\"{u}r Kernphysik, Saupfercheckweg 1, 69117 Heidelberg, Germany}
\affiliation{Institut f\"{u}r Theoretische Physik, Ruprecht-Karls-Universit\"{a}t Heidelberg, Philosophenweg 16, 69120 Heidelberg, Germany}
\author{J\"org Evers}
\affiliation{Max-Planck-Institut f\"{u}r Kernphysik, Saupfercheckweg 1, 69117 Heidelberg, Germany}

\date{\today}

\begin{abstract}
Light propagation through an ensemble of ultra-cold Rydberg atoms in electromagnetically induced transparency (EIT) configuration is studied. In strongly interacting Rydberg EIT media, non-linear optical effects lead to a non-trivial dependence of the degree of probe beam  attenuation on the medium density and on its initial intensity. We develop a Monte Carlo rate equation model that self-consistently includes the effect of the probe beam attenuation to investigate the steady state of the Rydberg medium driven by two laser fields. We compare our results to recent experimental data and to results of other state-of-the-art models for light propagation in Rydberg EIT-media. We find that for low probe field intensities, our results match the experimental data best if a density-dependent dephasing rate is included in the model. At higher probe intensities, our model deviates from other theoretical approaches, as it predicts a spectral asymmetry together with line broadening. These are likely due to off-resonant excitation channels, which however have not been observed in recent experiments. Atomic motion and coupling to additional Rydberg levels are discussed as possible origins for these deviations. 
\end{abstract}

\maketitle

\section{Introduction}

Electromagnetically induced transparency (EIT) in Rydberg gases has been the subject of intense studies both theoretically \cite{reslen2011, gorshkov2011,gorshkov2012, pritchard2012, sevincli2011,sevincli2011b, petrosyan2011, ates2011, petrosyan2012, yan2012, stanojevic2013, otterbach2013, tanasittikosol2011} and experimentally \cite{schempp2010, pritchard2010,pritchard2011, peyronel2012, dudin2012b, maxwell2013,hofmann2012} in the recent years. One motivation is to achieve strong interactions between photons by interfacing them with interacting states of matter. In particular, based on the excitation blockade \cite{lukin2001},  non-classical states of light can be prepared out of an initially classical driving field~\cite{peyronel2012, dudin2012b, maxwell2013, hofmann2012}. Possible applications include deterministic single photon sources, storage and retrieval of photons, as well as quantum gates based on photon-photon interactions. However, already the simulation of classical light propagating through a strongly interacting medium is a substantial theoretical challenge due to the high complexity of the underlying many-body physics. At the heart of this is the exponential complexity of the quantum many-body problem of interacting 3-level atoms and the non-linearity and non-locality of the propagation equations of the light related to the long range interactions. 

Various approaches using different approximations have been pursued to tackle light propagation through Rydberg EIT media. Sevin\c{c}li {\it et al.}\ \cite{sevincli2011} derived an analytical expression for the third order optical non-linearity based on the cluster expansion approach \cite{schempp2010}. This approach yields interesting results for moderate atomic densities but the cluster expansion is expected to break down at high densities \cite{ates2011}.
In the weak probe regime, where the probe field consists only of a few photons significant progress has been made recently \cite{gorshkov2012, peyronel2012}. However, for more than two photons in the probe field and imperfect EIT, numerical calculations become very demanding.
Petrosyan {\it et al.}\ \cite{petrosyan2011} developed a model including correlations in the light field. This model is based on coarse graining the atomic medium by introducing super-atoms.

All these approaches treat the atomic cloud as a continuous medium. Alternatively, the atoms can be treated individually as discrete objects. This has the advantage that the simulation can realistically model non-homogeneous trap geometries, large atom numbers and densities. However, models focusing on the atomic properties, such as inter-atomic correlations and other many-body effects, generally have the problem that the dimension of the Hilbert space grows exponentially with the number of atoms. This problem can be overcome by excluding states that are never populated due to the Rydberg blockade effect \cite{robicheaux2005, younge2009, gaerttner2012}. But this state reduction is not possible for non-Rydberg excited states, since they are not affected by strong interaction-induced level shifts.
If the driving is far off-resonant from the intermediate level, non-Rydberg excited states are never populated, and can be eliminated adiabatically.
In an EIT configuration, however, where both lasers are near-resonant with a low-lying intermediate excited state, this adiabatic elimination is not possible.
As a consequence, the state space truncation becomes ineffective. A further restriction arises because incoherent processes such as the spontaneous decay of the intermediate level are important in the EIT setting, which make a full master equation (ME) treatment necessary.

In order to overcome these difficulties, here, we use a model based on the rate equation (RE) ansatz developed by Ates {\it et al.}\ \cite{ates2007a,ates2007b,ates2011} and extended by Heeg {\it et al.}\ \cite{heeg2012} to calculate the steady state of a cloud of three-level atoms subject to coherent laser driving. In this model, interactions are included as level shifts only, making a classical Monte Carlo treatment possible. A strength of the RE model is that it enables one to obtain theoretical predictions over a broad parameter range. Since calculation times scale almost linearly with the atom number (compared to exponential in the case of the full ME), large atom numbers and densities in arbitrary geometries can be treated. Thus, calculations considering the actual experimental conditions become feasible, in parameter ranges inaccessible with ME or truncated Hilbert space models. One has to keep in mind, however, that the validity of the approximations entering into the RE model depends on the chosen parameters, and there are parameter conditions where the this approach is known to fail. One way of estimating the predictive power of RE based models are comparisons with other theoretical models, or benchmark calculations with more accurate models such as exact full ME simulations. But since it is the main motivation for using RE based models to access broader parameter space, the latter are usually only possible over a strongly restricted parameter range.

As a first example, we compare our results to recent experimental data obtained for resonant probe fields of low intensity~\cite{hofmann2012}. In the regime of weak probe laser fields, the probe beam is attenuated while traveling through the atomic cloud. For this, we extend the existing RE models to include absorption based on the propagation equations of classical light fields. These lead to spatially varying local probe fields experienced by the different atoms, and we solve the combined propagation equations and RE self-consistently.
We find that best agreement is achieved if in addition to the constant dephasing induced by the finite laser linewidth also a density-dependent dephasing is introduced. This additional dephasing could arise from motion-induced dephasing, and we find that the collision rates one obtains from a simple estimate based on kinetic gas theory are comparable to the relevant experimental time scales. Another effect that would also lead to density dependent enhancement of absorption is the coupling of the Rydberg state excited by the lasers to neighboring Rydberg levels.
We further study light propagation with off-resonant probe fields, and compare our results to those of other models \cite{petrosyan2011,sevincli2011}. We find that the models disagree at higher probe intensities, as the RE include resonant excitation channels at off-resonant laser driving which are not captured in the other super-atom based models. The resulting asymmetry in the spectra predicted by the RE, however, were not observed in recent experiments~\cite{pritchard2010,schempp2010}. A possible explanation for this discrepancy is that atomic motion could render the resonant excitation channels ineffective \cite{li2013}.

\begin{figure}[t]
  \centering
 \includegraphics[trim=1cm 1cm 1cm 0cm, width=\columnwidth]{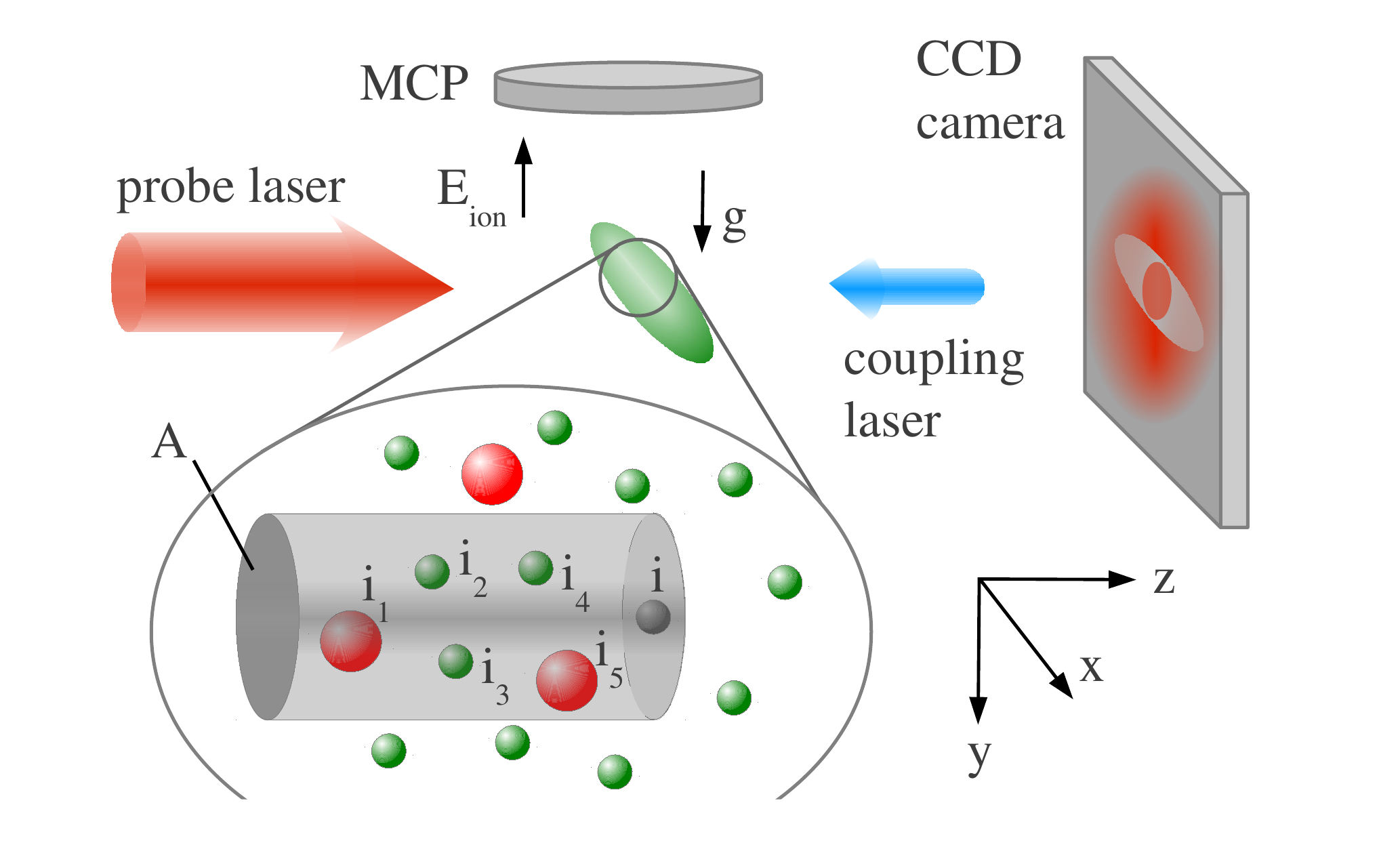}
 \caption{(Color online) Setup considered in the numerical calculations, adapted from~\cite{hofmann2012}. A cylindrical cloud of Rydberg atoms interacts with counter-propagating probe and coupling laser fields. We model the attenuation of the probe laser field to evaluate its intensity at the position of atom $i$ by considering an attenuation tube of transverse area $A$, as explained in the main text. Within this tube, large red spheres represent Rydberg excited atoms, while small green ones are atoms in non-Rydberg states at the time of evaluation. The internal states of the atoms in the tube determines the amount of attenuation. Next to the light absorption, we also calculate the number of Rydberg excitations, indicated by an ionizing field $E_{\text{ion}}$ and an ion detector (MCP). $g$ indicates gravitation also included as classical motion in our calculations.}
 \label{fig:setup}
\end{figure}

\section{Model description}
\label{sec:model}

\subsection{Monte Carlo rate equation model}

The RE model provides a way to calculate the steady state of a strongly interacting many-body system subject to lasers in EIT configuration that scales almost linear with the atom number as long as the Rydberg excited fraction is small. For our calculations, we mainly refer to Rydberg EIT experiments in the strong interaction regime as recently studied in Ref.~\cite{hofmann2012,hofmann2013}: The ground state $\ket{g}=\ket{5S_{1/2}}$ of $^{87}$Rb is coupled to an intermediate state $\ket{e}=\ket{5P_{3/2}}$ by the (weak) probe laser with Rabi frequency $\Omega_p$. The state $\ket{e}$ is coupled to the Rydberg state $\ket{R}=\ket{55S_{1/2}}$ by the (strong) coupling laser with Rabi frequency $\Omega_c$. The intermediate state $\ket{e}$ can spontaneously decay to the ground state with rate $\Gamma$, while the Rydberg state is long lived. The additional dephasings caused by the finite laser bandwidths lead to the total line widths $\gamma_{eg}$ and $\gamma_{gR}$ of the probe transition and the two photon transition, respectively. Two atoms that are in the Rydberg state show repulsive Van der Waals interaction with $C_6/2\pi=50\,$GHz$\mu$m$^6$.

The Hamiltonian of an ensemble of $N$ such atoms, in rotating wave approximation, reads ($\hbar=1$)
\begin{equation}
H=\sum_{i=1}^N \left[H_L^{(i)}+H_\Delta^{(i)}\right] + \sum_{i<j}\frac{C_6 \ket{R_i R_j}\bra{R_i R_j}}{|\mathbf{r}_i-\mathbf{r}_j|^6} 
 \label{eq:Hamiltonian}
\end{equation}
where
\begin{align}
H_L^{(i)}=\Omega_p/2 \ket{g_i}\bra{e_i} + \Omega_c/2 \ket{e_i}\bra{R_i} + h.c.
\end{align}
describes the coupling of the atoms to the laser fields and 
\begin{align}
  H_\Delta^{(i)}=-\Delta_1\ket{e_i}\bra{e_i} - (\Delta_1 + \Delta)\ket{R_i}\bra{R_i}
\end{align}
accounts for the detuning from the one and two photon resonance.
Incoherent processes can be included as Lindblad terms $\mathcal{L}[\rho]$~\cite{agarwal74,ficek05,kiffner10,fleischhauer2005} leading to the ME for the density matrix
\begin{equation}
 \dot{\rho}=-i[H,\rho]+\mathcal{L}[\rho]\,.
\end{equation}
For a single atom ($N=1$) one can transform the ME into a set of RE for the populations of the atomic levels by adiabatically eliminating the coherences ($\dot{\rho}_{ij}=0$ for $i\neq j$) \cite{ates2007a}. For the many-body case one can intuitively generalize this to a RE for the populations of the product states $\ket{\boldsymbol{\sigma}} = \ket{\sigma_1,\sigma_2,\ldots ,\sigma_N}$, where $\sigma_i\in\{g,e,R\}$. We employ a Monte Carlo technique for the solution of the many body RE, that is, starting in the global ground state $\ket{g,g,\ldots ,g}$ we perform a random walk through the configuration space of states $\ket{\boldsymbol{\sigma}}$~\cite{heeg2012} and average over many such trajectories, ensuring the convergence to a global steady state. The Hamiltonian $H$ couples two such many body states only if they differ in exactly the state of one atom. Therefore, it is sufficient to randomly pick one atom in each Monte Carlo step and determine the probability (jump rate) with which its state is changed. 
In order to calculate these rates, a mean-field-like approximation is required: The interaction between atoms in the Rydberg state is incorporated merely as a shift of the Rydberg level of the considered atom $\Delta\ind{int}^{(i)}=\sum_{j\neq i}^{\prime} V_{ij}$, where the sum only runs over atoms that are currently in the Rydberg state. $\Delta\ind{int}^{(i)}$ enters as an additional detuning into the ME of atom $i$, i.~e., the detuning of the coupling laser for atom $i$ is modified according to $\Delta\rightarrow\Delta^{(i)}=\Delta-\Delta\ind{int}^{(i)}$. 
This generalization to the many-body case is not unique, but it can be shown to capture many relevant features of the many-body system~\cite{gaerttner2013c}. As the involved approximations may fail depending on the chosen parameters, in the following, we also discuss benchmark comparisons of our numerical results to exact full ME simulations for few particles.

\subsection{Including propagation effects}

We now discuss how the attenuation of the probe beam can be included in the RE model.
Classical light propagating through an atomic medium with electric susceptibility $\chi=\mathrm{Im}(\chi_{eg})$ and thickness $L$ is damped exponentially  
\begin{align}
\Omega_p(L) = \Omega_p(0) e^{-\chi k L/2}\,,
\end{align}
where k is the wave vector of the light. For resonant probe fields, dispersion and transverse beam dynamics can be neglected~\cite{sevincli2011}. In terms of atomic properties $\chi$ is given as
\begin{equation}
\label{eq:chi1}
 \chi = \frac{2|\mu_{eg}|^2 n_0}{\epsilon_0 \hbar \Omega_{p}}\mathrm{Im(\rho_{ge})} = \frac{3\lambda^2 n_0 \Gamma^2}{2\pi k \Omega_{p}^2}\rho_{ee}
\end{equation}
where $\mu_{eg}$ is the dipole matrix element of the probe transition, $n_0$ the atomic density, $\lambda=2\pi/k$ the probe wavelength, and $\Gamma$ the spontaneous decay rate from $\ket{e}$ to $\ket{g}$.

In order to include the propagation effect in the Monte Carlo simulation, we have to calculate the local probe Rabi frequency that a certain atom $i$ experiences. For this, we define a cylindrical volume (tube) of cross section $A$ located around atom $i$ and extending into the opposite direction of the probe light propagation, see Fig.~\ref{fig:setup}. All atoms inside this tubes contribute to the attenuation of the probe beam before it reaches atom $i$. The attenuation is calculated recursively, starting at the first atom in the tube ($i_1$), that experiences the full probe laser power corresponding to the Rabi frequency $\Omega_p^{(0)}$. Using $\Omega_p^{(0)}$ we calculate the steady state value of $\rho_{ee}^{(i_1)}$ for the current configuration $\ket{\boldsymbol{\sigma}}$ and use this to determine the Rabi frequency behind atom $i_1$ as 
\begin{equation}
 \Omega_p^{(i_1)} = \Omega_p^{(0)} \exp\left[-\frac{3\lambda^2\Gamma^2\rho_{ee}^{(i_1)}}{4\pi A (\Omega_p^{(0)})^2}\right].
\end{equation}
Using $\Omega_p^{(i_1)}$ this procedure is repeated with the next atom $i_2$ in the tube and so on until atom i is reached. 
The local Rabi frequency $\Omega_p^{(i)}$ is then used to determine the steady states and thus the jump probabilities for atom $i$ and to update its state. This procedure is repeated until the global observables converge. Additionally we average over many random Monte Carlo samples of atom positions.

In our numerical routines the recursive calculation of $\Omega_p$ is not required in every step. Instead, the values of the local susceptibility and Rabi frequency are stored and reused. They only have to be updated, when an atom jumps into the Rydberg state or out of the Rydberg state, since in this case the interaction shifts of all other atoms change.

The only parameter that we can choose freely is the tube cross section $A$. We found that the results are independent of the exact choice of $A$ as long as two criteria are fulfilled: $A$ must be large enough to obtain $N\ind{tube}\gg 1$ atoms per tube on average, and it must be small enough, such that the atomic density does not vary much over the tube diameter. When simulating samples of varying density, we choose $A$ such that the average $N\ind{tube}$ is the same for all densities.

\section{Results}

\subsection{Density dependence on resonance}
\label{sec:comp_weide}

\begin{figure}[t]
  \centering
 \includegraphics[width=\columnwidth]{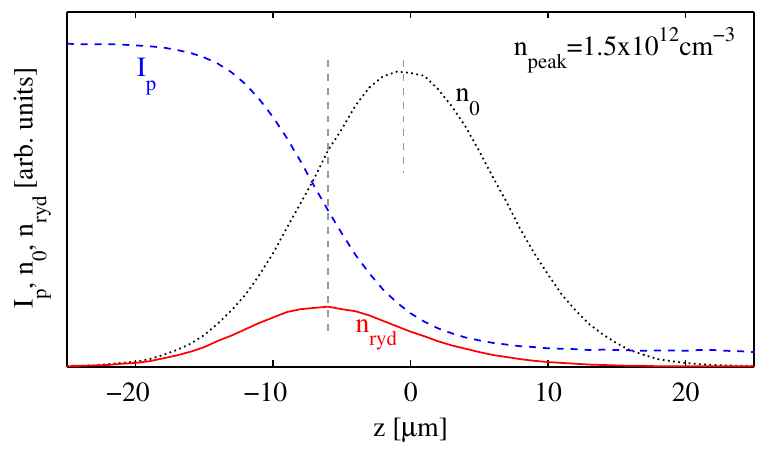}
 \caption{(Color online) Probe beam intensity $I_p$ (dashed blue), atomic density $n_0$ (dotted black), and density of Rydberg excitations $n\ind{ryd}$ (solid red) along the propagation direction of the probe beam. The Rydberg density has been amplified by a factor of 500 with respect to the atomic density. The peak value of $n_0$ is $1.5\times10^{12}\,$cm$^{-3}$. Parameters are $\Omega_p^{(0)}/2\pi=0.235\,$MHz, $\Omega_c/2\pi=5.1\,$MHz, $\Gamma/2\pi=6.1\,$MHz, $\gamma_{eg}/2\pi=6.4\,$MHz, $\gamma_{gR}/2\pi=1.7\,$MHz, and $C_6/2\pi=50\,$GHz$\,\mu$m$^6$.}
 \label{fig:z_dep}
\end{figure}

In the first part, we consider the setup in Fig.~\ref{fig:setup}, and compare our theoretical predictions to corresponding experimental data reported in~\cite{hofmann2012}. As sketched in Fig.~\ref{fig:setup}, a small ensemble of $^{87}$Rb atoms is illuminated by counter-propagating probe and coupling lasers, where the coupling laser is focused to a small spot. We calculate the absorption image of the could as well as the number of produced Rydberg excitations, as a function of the atomic density of the Gaussian-shaped cloud. The laser parameters used throughout this section are given in the caption of Fig.~\ref{fig:z_dep}. 

\begin{figure}[t]
  \centering
 \includegraphics[width=\columnwidth]{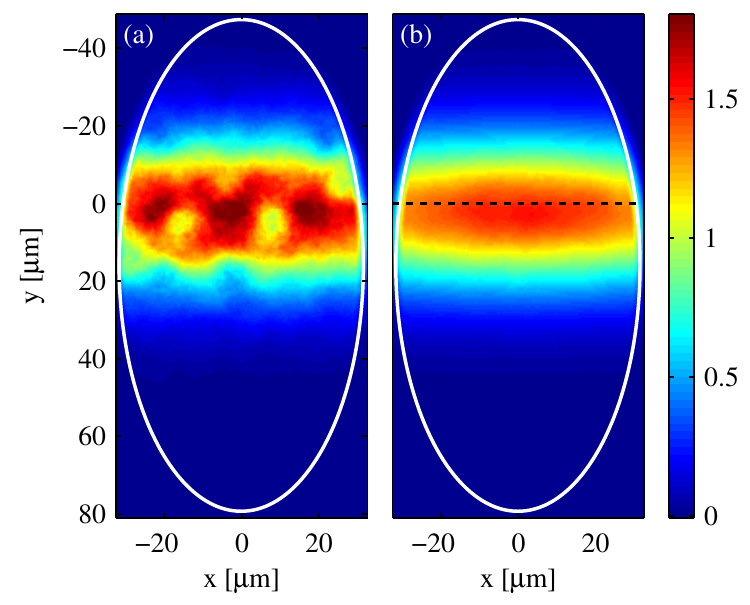}
 \caption{(Color online) Simulated absorption images of the atomic cloud. We plot the relative difference between the transmitted probe Rabi-frequency and the respective two level response, $(\Omega_p-\Omega_p^{(2L)})/\Omega_p^{(2L)}$. (a) shows a snapshot of a single Monte Carlo trajectory. We observe two prominent structures near the center stemming from Rydberg excitations that cause enhanced absorption in their vicinity. The peak density is chosen as $2.8\times10^{11}\,$cm$^{-3}$ in this figure. The white ellipse marks the border of the coupling laser spot. (b) Average over 500 Monte Carlo samples. The dashed line marks the position of the original center of the cloud before falling under gravity.}
 \label{fig:EITspot}
\end{figure}

Figure \ref{fig:z_dep} shows how the probe beam is attenuated while propagating through the atomic cloud. The higher the atomic density, the faster the probe intensity drops. Therefore the maximum of the Rydberg density does not coincide with the maximum of the atomic density. This is indicated by the dashed vertical lines in Fig.~\ref{fig:z_dep}. The next quantity of interest is the transmitted probe intensity relative to the respective intensity observed in the two-level medium obtained in the absence of the coupling beam (without EIT). Figure \ref{fig:EITspot} shows the distribution of this relative intensity in a section transverse to the beam propagation direction. In (a), a single Monte Carlo trajectory is shown. The noise is due to fluctuations in the local atomic density. The two dips close to the trap center are signatures of Rydberg excitations reducing the transmission in their vicinity. Such images cannot be obtained easily with current state-of-the-art experiments since the exposition time required to obtain an absorption image of sufficient signal to noise ratio is long on the time scale of the excitation dynamics. Thus excitations will vanish and reappear at other positions while the image is acquired making the spatially resolved detection of Rydberg excitations impossible. To overcome this difficulty, alternative imaging schemes have been proposed \cite{guenter2012,olmos2011}.

In typical experiments, a time-integrated transmission signal is recorded, which is in addition averaged over several repetitions of the experiment. This procedure is mimicked in our Monte Carlo simulation by averaging over several Monte Carlo trajectories and several realizations of randomly chosen atom positions. Such an averaging results in a transmission pattern as shown in Fig.~\ref{fig:EITspot}(b) which can be compared directly to camera images obtained in the experiment reported in~\cite{hofmann2012}.

\begin{figure}[t]
  \centering
 \includegraphics[width=\columnwidth]{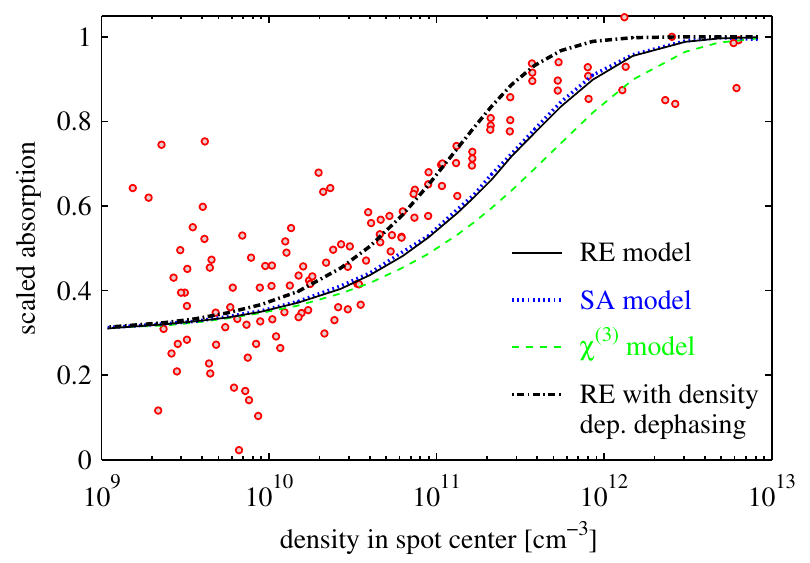}
 \caption{(Color online) Density dependence of scaled absorption. Red open circles: experimental data~\cite{hofmann2012}, solid black line: RE model, dotted blue line: super atom model \cite{petrosyan2011}, dashed green line: Calculations using the third order susceptibility from \cite{sevincli2011}. The dot-dashed line is obtained by including additional atomic motion induced dephasings. The experimental data was acquired over an exposure time of $100\,\mu$s, much longer than the excitation time $2\,\mu$s for the data in Fig.~\ref{fig:Nryd_dep}.}
 \label{fig:abs_dep}
\end{figure}

We simulated the probe intensity behind the cloud ($z=\infty$) in the center of the excitation region ($x=y=0$). The results for the EIT-absorption are divided by the absorption obtained with the coupling laser switched off in order to eliminate trivial density dependences. In the low and high density limit the results (see solid black line in Fig.~\ref{fig:abs_dep}) agree well with the experimental data from~\cite{hofmann2012} (red open circles in Fig.~\ref{fig:abs_dep}). However, at intermediate densities, the experimentally observed scaled absorption is clearly underestimated by the RE model. In order to understand this discrepancy, we inspect the four major approximations that enter into our calculations. These are, first, the inclusion of interactions as mere level shifts, which is the main approximation of the RE model, second, the classical treatment of the light propagation, third, the frozen gas approximation, and fourth, the assumption of a single Rydberg level. We note that the simulations of the scaled absorption have no adjustable parameters. In~\cite{hofmann2012}, all experimental parameters have been determined in independent measurements.

\begin{figure}[t]
  \centering
 \includegraphics[width=\columnwidth]{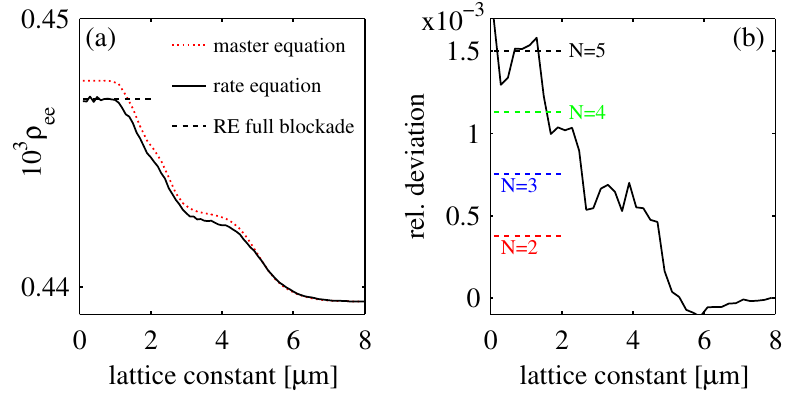}
 \caption{(Color online) Comparison between rate equation and master equation for few atoms in a lattice configuration. The parameters are the same as in Fig.~\ref{fig:abs_dep}. The atoms are arranged in a regular lattice with varying lattice spacing. The limit of small lattice spacing corresponds to full blockade, while for large spacing the atoms are non-interacting. (a) Intermediate state population $\rho_{ee}$ and $\mathrm{Im}[\rho_{ge}]\Gamma/\Omega_p$ as a function of lattice spacing for $N=5$ atoms. We additionally show the analytical solution of the rate equation for a fully blockaded ensemble  as dashed line. (b) Relative difference between rate equation and master equation. Solid line: $N=5$, dashed lines:  deviation in the full blockade case for other atom numbers.}
 \label{fig:benchmark_weide}
\end{figure}

\begin{figure}[t]
 \centering
 \includegraphics[width=\columnwidth]{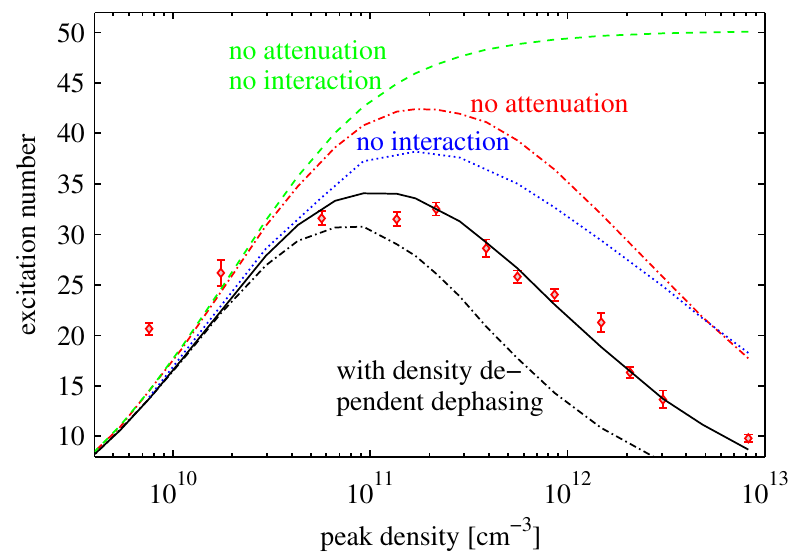}
 \caption{(Color online) Number of excited atoms as a function of cloud density. The detector efficiency $\eta$ and the semi-major axis of the excitation spot used in the simulations are $0.4$ and $65\,\mu$m, respectively~\cite{hofmann2012}. Red dots show experimental data~\cite{hofmann2012}. In addition to the full simulation results (black solid line), also curves with probe beam attenuation and/or inter-atomic interaction switched off are shown for comparison. As explained in the main text, the dash-dotted curve in addition includes a density-dependent dephasing, which is not expected to occur at the short exposure time of $2\mu$s at which the Rydberg excitations were recorded. }
 \label{fig:Nryd_dep}
\end{figure}

In order to check whether the local medium response is reproduced correctly by the RE model, we benchmark it by comparing it to full ME simulations.  For this, we recall that in the RE model with probe absorption, the local susceptibility is calculated from the intermediate state population using $\mathrm{Im}[\rho_{ge}] = \rho_{ee}\Gamma/\Omega_p$. We therefore compare the intermediate state population obtained from the RE model to $\mathrm{Im}[\rho_{ge}]\Omega_p/\Gamma$ from full ME calculations. Due to the exponential growth of the state space with the number of 3-level atoms, the ME simulations are restricted to only few atoms. The atoms are placed in a regular chain and the distance between neighboring atoms is varied. Small lattice spacing corresponds to high density, while for large lattice spacing the non-interacting regime is approached.  The results are shown in Fig.~\ref{fig:benchmark_weide} for up to 5 atoms.  The parameters are as in Figs.~\ref{fig:abs_dep} and \ref{fig:Nryd_dep}. We find that the probe beam absorption is underestimated systematically by the RE model, and the deviation to the ME result increases with density. The deviations, however, are only on the order of $10^{-3}$ for five fully blockaded atoms, which corresponds to a density of about $10^{10}\,$cm$^{-3}$ (solving $N_b=n_0V_{b}=5$ for $n_0$ with $V_b=4\pi r_b^3 /3$ and $r_b=5\,\mu$m). Higher densities are not accessible for the ME, as then there would be more atoms per blockade radius than included in the simulation. At density $10^{11}\,$cm$^{-3}$, where the deviation in scaled absorption between theory and experiment is largest, there are approximately $50$ atoms per blockade radius, inaccessible to ME treatments.

For up to 5 atoms, the deviation approximately increases linearly with the number of blockaded atoms. Naively extrapolating this linear dependence to higher densities would lead to a deviation in $\rho_{ee}$ of order of $1\%$ at a density of $10^{11}\,$cm$^{-3}$. The relative differences in the Rydberg population are of the same order. This would not be sufficient to explain the deviations from the experimental data of $\gtrsim 10\%$. Obviously, this linear extrapolation is expected to break down at higher densities.
But in related calculations, as expected  we found that the RE model generally  performs better as dephasing rates increase compared to the coherent drive. This was also pointed out in \cite{petrosyan2012b} for the case of two-level atoms~\cite{gaerttner2013c}. In our present calculations, the dephasing rates are quite large compared to the probe Rabi frequency, as $\sqrt{N}\Omega_p^{(0)}$ only exceeds $\gamma_{gR}$ starting from  $N\approx 50$. This explains the good performance of the RE model in the lower density regime, and suggests that its validity range extends into the region of substantial deviation between theory and experiment in Fig.~\ref{fig:abs_dep}.

Since a direct benchmark of the RE results to corresponding ME results is possible only over a restricted density range, an alternative strategy to investigate the validity of the RE approach is to compare theory and experiment for other observables in the parameter range inaccessible to ME treatments. In particular, the RE model also gives access to the Rydberg excitations. The predicted number of excitations agrees well with experimental values of Ref.~\cite{hofmann2012} over the entire density range, see Fig.~\ref{fig:Nryd_dep}. Here, we adjusted two parameters that were not determined from independent measurements. Namely, the semi-major axis of the coupling laser spot was found to be $65\,\mu$m, and the detection efficiency of the MCP was found to be $\eta=0.4$, in accordance with Ref.~\cite{hofmann2012}. Note that this data was taken after an excitation of $2\mu$s, such that motional dephasing is not expected  to be relevant here. We have added the results for the excitation number that we obtain if we exclude attenuation and interaction effects (green dashed line in Fig.~\ref{fig:Nryd_dep}). The obtained number of excitations is given by $f_0N$, where $f_0$ is the single atom excitation probability. Additionally we simulated the system excluding interactions but including attenuation effects and vice versa. The strong deviations from the experimental data at high densities in both cases show that both, attenuation of the probe beam and interaction between the atoms, have a significant impact on the number of produced Rydberg excitations. This means that including the probe beam attenuation self-consistently in the RE model is indispensable for the simulation of Rydberg EIT in a dense gas. 
The good agreement of the Rydberg population with the experimental data is a further indication that the comparison between RE theory and experiment in Fig.~\ref{fig:Nryd_dep} is meaningful. 

In order to address possible issues with the light propagation, we compare our results to a model proposed by Petrosyan {\it et al.}~\cite{petrosyan2011}. This work makes use of a simple super-atom model for the atom dynamics and focuses on the propagated light which is characterized via coupled propagation equations for the intensity and the correlation function of the probe light. This way, correlations in the light field going beyond the classical treatment in our approach can be included. The model describes light propagation through a one-dimensional array of super-atoms with diameter $2r_b$. The blockade radius $r_b$ is defined by equating the EIT-width to $C_6/r_b^6$. Interactions between super-atoms are included as a small mean field shift appearing in the susceptibility, which will be discussed in more detail in Sec.~\ref{sec:delta_scan}. We extended the original model by replacing the EIT width $w=|\Omega_c|^2/\gamma_{eg}$ by $\gamma_{gR}+|\Omega_c|^2/\gamma_{eg}$ due to the larger dephasing rates in our setup, such that the contribution $\gamma_{gR}$ can not be neglected. Furthermore we include spatially varying densities, i.e., the number of atoms per super-atom $n_{SA}$ becomes spatially dependent. With the above extensions, we obtain very good agreement for the properties of the propagated light between the two models. However, we found that for our parameters, the simulation results remain unchanged if the photon statistics is forced to remain classical in the extended model of Petrosyan {\it et al.}. For this, we set the $g^{(2)}$ of the light field to one. 
This indicates that for the parameters of this experiment, the non-classical character of the light does not influence the total absorption.

As a further cross check for our model, in Fig.~\ref{fig:abs_dep}(a) we show the scaled absorption obtained including the third order non-linear absorption calculated in \cite{sevincli2011}. But this model deviates stronger from the experimental data in the relevant density regime. One reason for this could be that  the original assumption of neglecting the transverse beam profile exploited in \cite{sevincli2011} to derive an analytic expression for the nonlinear susceptibility is not satisfied for the present parameters, since the density varies rapidly perpendicular to the propagation direction. Moreover, this model is based on a truncation in the correlation order at the two particle level, and is thus expected to fail at high densities, where higher order correlations become crucial.

The third key assumption is the frozen gas approximation. Higher absorption could be caused by atomic motion induced dephasing. In the experimental situation under discussion, a thermal cloud of atoms at $T=5\,\mu$K is considered. The average speed of an atom is thus $v=\sqrt{8kT/\pi m}=0.035\,$m/s. This means that within the excitation time of $100\,\mu$s an atom typically moves across a distance of $3.5\,\mu$m. As a consequence, in a binary picture, an atom that is initially unblockaded with respect to second atom, can move towards the second atom within the excitation time and undergo a collision that entangles the internal with the motional degrees of freedom and therefore leads to decoherence of the internal dynamics. Estimating the collision rate from classical kinetic gas theory, we obtain $n\ind{coll}=\sigma v n_0\approx 1\,\mu$s$^{-1}$ at a density of $n_0=10^{11}\,$cm$^{-3}$. 
Here, the scattering cross section $\sigma=\pi r_t^2$ is determined by estimating the classical turning point from $m v^2/2=\hbar C_6/r_t^6$. This means that after an excitation time of $100\,\mu$s, essentially all atoms would have undergone several such collisions. 
From this estimate, one would  expect a motion-induced additional dephasing of the the Rydberg level which is proportional to the atomic density. We test our hypothesis of an additional dephasing proportional to the atomic density by adding a dephasing rate  $\Gamma_{R,mot}/2\pi=\alpha n_0$ to our model. The result is the dot-dashed curve in Fig.~\ref{fig:abs_dep}, which shows good agreement with the experimental data. This curve was obtained with $\alpha = 1.2\times 10^{-11}\,$MHz\,cm$^3$, which is of the same order of magnitude as the estimated collision rate $n\ind{coll}\approx 10^{-11}\,$MHz\,cm$^{3}n_0$. For the given value of $\alpha$, the density-dependent dephasing exceeds the constant laser-induced dephasing for densities larger than approximately $1.5\times 10^{11}\,$cm$^{-3}$.
It should be noted, however, that a quantitative estimate of such a dephasing rate  would require a study of the underlying mechanism of dephasing collisions on the microscopic scale which is beyond the scope of this work. 

We also studied the effect of the density dependent dephasing on the number of Rydberg excitations shown  in Fig.~\ref{fig:Nryd_dep} (dot-dashed line). We find that with the additional density-dependent dephasing rate, the Rydberg excitations are severely underestimated at high densities. Since the Rydberg excitations were recorded after a short exposure time of $2\mu$s at which motional effects are not expected to be significant, we interpret this result as a further indication that the deviations in absorption in Fig.~\ref{fig:abs_dep} are caused by a mechanism that is only relevant for long excitation times, consistent with motion-induced dephasing. 

Density dependent dephasing effects have recently been studied in hot atomic vapors  \cite{baluktsian2013} (see also \cite{raitzsch2009}). The setup in this experiment is different from ours as the excitation lasers are far detuned from the intermediate level. Nevertheless, a linear dependence of the dephasing on the atomic density was found in this work as well. Additionally, in~\cite{baluktsian2013}, the motional dephasing was found to be proportional to the Rydberg population $f_R$. 
While we have employed a motional dephasing that is independent of $f_R$ in our calculations, we note that we checked that an additional dephasing term proportional to the Rydberg density $f_R n_0$ instead of the atomic density $n_0$ alone would also lead to good agreement between theory and experiment in our case.

Finally, we investigate the truncation of the level space to three-level atoms. In Ref.~\cite{hofmann2012} signatures for transfer of Rydberg excitations to adjacent states have been observed. Such excitations would be excluded from the laser dynamics, and thus effectively become meta-stable. This effect would lead to an increased number of Rydberg excitations at long excitation times and could therefore enhance absorption. The significance of additional Rydberg excitations is expected to depend on the number of particles per blockade volume and thus on the atomic density. Opposite to the motional dephasing, this effect would result in slowly in creasing number of Rydberg excitations and could be checked for experimentally by state selective ionization. Excitation of neighboring Rydberg levels at long excitation times has also been observed in \cite{peyronel2012}.

\subsection{Dependence on probe field detuning}
\label{sec:delta_scan}

So far, we have only considered resonant probe and coupling beams. Next, we study the dependence of the transmission through an elongated cloud of length $L=1.3\,$mm and constant density $n_0=1.2\times 10^{10}\,$cm$^{-3}$ on the probe field detuning. The laser parameters are as in Refs.~\cite{pritchard2010,petrosyan2011}. Dephasings are smaller compared to Ref.~\cite{hofmann2012} and $C_6$ is larger (a $\ket{60s}$ state with $C_6/2\pi=140\,$GHz\,$\mu$m$^6$ is used). In the super-atom model of Ref.~\cite{petrosyan2011}, the correlation function of the light field was included to account for the emergence of non-classical states of light.

\begin{figure}[t]
  \centering
 \includegraphics[width=\columnwidth]{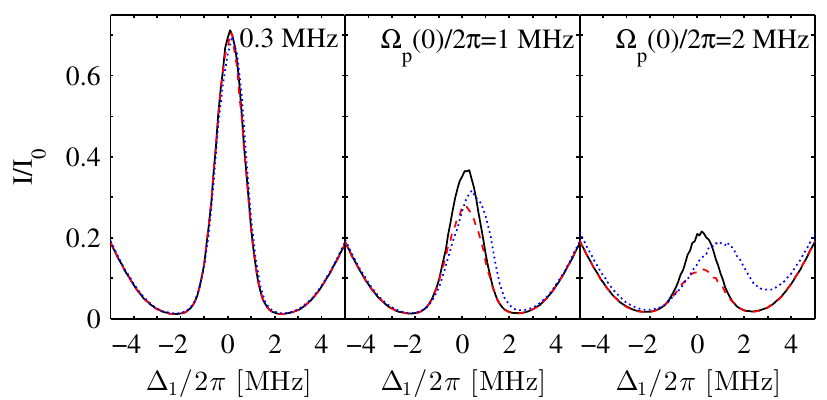}
 \caption{(Color online) Transmission through an elongated cloud ($L=1.3\,$mm) of density $1.2\times 10^{10}\,$cm$^{-3}$ as a function of probe detuning and intensity. Remaining parameters are $C_6/2\pi=140\,$GHz$\,\mu$m$^6$, $\Omega_c/2\pi=4.5\,$MHz, $\gamma_{eg}/2\pi=6.1\,$MHz, $\gamma_{gR}/2\pi=0.1\,$MHz and $\Delta/2\pi=-0.1\,$MHz as in Ref.~\cite{pritchard2010}. Solid black line: super-atom model, dotted blue line: RE model, dashed red line: super atom model with $g^{(2)}=1$.}
 \label{fig:delta_dep}
\end{figure}

Scanning the probe detuning $\Delta_1$ for various initial probe Rabi frequencies $\Omega_p(0)$, we obtain the transmission curves depicted in Fig.~\ref{fig:delta_dep}. For low probe intensity the models agree well. In this case $g^{(2)}$ does not deviate much from unity. As the probe intensity is increased, the transmission on resonance decreases showing the non-linearity of the process. The transmission obtained from the RE model shows a clear shift and broadening of the EIT resonance while the super-atom model does not. If $g^{(2)}$ is set to unity in the super atom model, the resulting shift and asymmetry is still small, while the main effect is a decrease of transmission near resonance. The asymmetry observed in the RE results is due to higher-order resonant excitation channels. If the interaction shift cancels the detuning, Rydberg excitation is enhanced (anti-blockade) which leads to smaller $\rho_{ee}$ and thus reduces absorption. As this only happens for positive detunings, the curve becomes asymmetric. 
The asymmetry is not present in the super-atom model since here interactions between different super-atoms are only included as a small mean field shift in the EIT-absorption. This shift is indeed negligible for the parameters studied here and does not account for the anti-blockade.

\begin{figure}[t]
  \centering
 \includegraphics[width=\columnwidth]{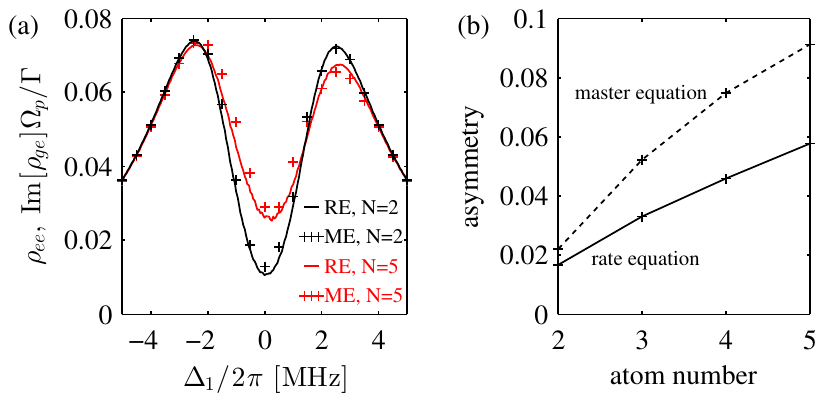}
 \caption{Comparison between RE and master equation for few atoms in a spherical trap. The parameters are comparable to the ones of Fig.~\ref{fig:delta_dep}(c). The atoms are placed randomly in a spherical trap of radius $5.5\,\mu$m. (a) Intermediate state population and rescaled imaginary part of $ge$-coherence (for RE and ME, respectively) as a function of probe laser detuning and for two different atom numbers (densities). (b) Asymmetry of the Rydberg population as a function of atom number.}
 \label{fig:benchmark_fleisch}
\end{figure}

Nevertheless, the asymmetry predicted by the RE model was not observed in related experiments~\cite{schempp2010,pritchard2010}, which invites a further investigation.
For this, we next show that this asymmetry is not an artifact of the RE model, but is indeed underestimated by it, by comparing to exact ME calculation with few atoms. Fig.~\ref{fig:benchmark_fleisch} shows the result of a simulation with $2$ to $5$ atoms in a spherical trap with random position sampling. $N=5$ atoms corresponds to a density of $n_0=7\times 10^9\,$cm$^{-3}$. The remaining parameters are the same as in Fig.~\ref{fig:delta_dep}(c), except that we ignore the small detuning of the coupling laser ($\Delta=-0.1\times 2\pi\,$MHz), in order not to bias our asymmetry parameter by this small shift. Note that the overall shape of the curve is unchanged if we include this detuning. We observe that while for $N=2$ atoms the asymmetry is still rather small, it becomes increasingly pronounced at larger densities. We also found that increasing the system size holding the density constant renders the asymmetry even more pronounced. In Fig.~\ref{fig:benchmark_fleisch}(b) we quantitatively analyze the asymmetry by calculating the difference of the integral over the blue detuned side ($\Delta_1>0$) and the red detuned side ($\Delta_1<0$), normalized by the integral over the full range of $(-5 \leq \Delta_1 \leq  5) \times 2\pi\,$MHz. We observe that the asymmetry grows approximately linearly with the atom number (density) and is underestimated by the RE model, which we attribute to the fact that higher order resonant processes relying on higher-order atom correlations are not accounted for \cite{heeg2012, gaerttner2013b}. The large relative differences between ME and RE are due to the fact that the asymmetry parameter is very sensitive already to small deviations in the transmission spectra. But they also show that the predictions of the RE model can not always be trusted, and that the validity also depends strongly on the chosen observable.

This asymmetry is not present in the super-atom model because interactions between different super-atoms are only included as a small mean field shifts in the EIT-absorption which cannot account for an anti-blockade. But the physical reason why this asymmetry is not observed in experiment \cite{schempp2010,pritchard2010} must be different. One candidate are again atomic motion and effects beyond the frozen gas approximation. After a pair of atoms is excited resonantly, the atoms start repelling each other as they feel the repulsive force induced by the Rydberg-Rydberg interactions, thereby moving out of the pair excitation resonance. This effect can render resonant excitation processes inefficient for long exposure times. To estimate the relevance of this effect, we consider the case of $\Delta/2\pi=1\,$MHz. Two atoms can be excited resonantly if they are at a distance $r\ind{res}=[C_6/(2\Delta)]^{1/6}=6.4\,\mu$m. Assuming that both atoms get excited initially and calculating the classical trajectory on which the atoms move apart one obtains that after $10\,\mu$s the interatomic distance has increased by about $1\,\mu$m and the atoms have taken up a relative velocity of $0.13\,\mu$m$/\mu$s. Thus, they have moved out of the pair resonance, such that the double excitation probability decreases again, and they have received a momentum kick well above the mean thermal momenta at cryogenic temperatures. Thus the effect of resonant processes is rather a heating of the gas than an enhancement of the Rydberg population if excitation times are too long. These mechanisms have been studied recently in microtraps and optical lattice setups, concluding that motional effects can inhibit resonant pair excitation \cite{li2013}. Recalling that the data of Ref.~\cite{pritchard2010} was taken by scanning $\Delta/2\pi$ from $-20\,$MHz to $20\,$MHz in $500\,\mu$s it becomes clear that such effects should play a role, possibly enhanced by the dynamic frequency sweep. We note that for the case of attractive interactions it was found that the transmission spectrum strongly depends on the direction of the detuning scan, indicating that mechanical effects come into play \cite{pritchard2011}. Mechanical effects playing a role in this context have also been mentioned in Ref.~\cite{sevincli2011b}.
Next to motional effects, there are other possible reasons for the absence of an asymmetry in the experimental results of Ref.~\cite{pritchard2010}. For example, light propagation effects beyond pure absorption could play a role~\cite{peyronel2012}. We further note that by reducing the atomic density, Pritchard {\it et al.} did obtain an asymmetric transmission profile that matched very well the results of a three-atom master equation calculation, c.f.\ Fig.~4 in Ref.~\cite{pritchard2010}.

\section{Summary and discussion}
\label{sec:summary}

We have introduced an extended RE model including the attenuation of the probe beam, which is indispensable in the weak probe and high density regime. We applied our model to two different experimental situations: First, we simulate transmission of a weak probe beam through an atomic cloud at EIT resonance as a function of atomic density. Here we find good agreement with experimental results and other models was found for resonant laser driving in a large range of atomic densities. At high density and for experiments with long excitation times, we find that our model underestimates the probe absorption.  As potential origins of this discrepancy we discussed a motion induced density dependent dephasing and excitation of additional metastable Rydberg levels not coupled to a rapidly decaying state by the lasers. Second, we studied the dependence of the probe transmission on the single photon detuning and probe intensity at relatively low atomic density. We find that at low probe intensities, our model agrees well with the experimental data. But towards higher probe intensities, our model predicts a shift and broadening of the EIT resonance that is much stronger than observed experimentally. At the low density considered in the experiment, dephasing caused by collisional effects is expected to be small. However, the atomic motion can have another effect. Mechanical forces between resonantly excited pairs of atoms lead to a repulsion between them, which can render resonant excitation processes ineffective at long excitation times. Therefore, the results in both considered experimental settings suggest possible effects beyond the frozen gas approximation, and motivate further theoretical modeling and experimental studies on the validity of this approximation.

\begin{acknowledgments}
We thank  S.\ Whitlock, M. Robert-de-Saint-Vincent, G.\ G\"unter, C.\ S.\ Hofmann, H.\ Schempp, and M.\ Weidem\"uller for providing original experimental data and for insightful discussions. We thank K.\ P.\ Heeg for discussions and for work on the theoretical models. This work was supported by University of Heidelberg (Center for Quantum Dynamics, LGFG).
\end{acknowledgments}

\end{document}